\documentclass{WileyMSP-template}
\usepackage{amsmath}
\usepackage{color}

\begin{document}

\pagestyle{fancy}

\title{Dynamical Phase Transitions in Open Quantum Walks}

\maketitle


\author{Stefano Longhi}



\begin{affiliations}
Prof. Stefano Longhi\\
Dipartimento di Fisica, Politecnico di Milano,Piazza L. da Vinci 32, I-20133 Milano, Italy\\
\& IFISC (UIB-CSIC), Instituto de Fisica Interdisciplinar y Sistemas Complejos, E-07122 Palma de Mallorca, Spain \\
Email Address: stefano.longhi@polimi.it

\end{affiliations}



\begin{abstract}
Dynamical phase transitions in the relaxation behavior of stochastic quantum walks are investigated, focusing on systems where coherent unitary evolution is periodically interrupted by dephasing. This interplay leads to a classicalization of the dynamics, effectively described by non-equilibrium Markovian processes that can violate detailed balance. As a result, such systems exhibit a richer and more complex spectral structure than their equilibrium counterparts. Extending recent insights from classical Markov dynamics [G. Teza {\it et al.}, Phys. Rev. Lett. {\bf 130}, 207103 (2023)], we demonstrate that these quantum-classical hybrid systems can host not only first-order dynamical phase transitions -- characterized by eigenvalue crossings -- but also second-order transitions marked by the coalescence of eigenvalues and eigenvectors at exceptional points. We analyze two paradigmatic models: a quantum walk on a ring under gauge fields and a walk on a finite line with internal degrees of freedom, both exhibiting distinct mechanisms for breaking detailed balance. These findings reveal a novel class of critical behavior in open quantum systems, where decoherence-induced classicalization enables access to non-Hermitian spectral phenomena. Beyond their fundamental interest, our results offer promising implications for quantum technologies, including quantum simulation, error mitigation, and the engineering of controllable non-equilibrium quantum states.
\end{abstract}

\section{Introduction}

Phase transitions have long been a cornerstone of statistical physics, traditionally associated with abrupt changes in the equilibrium properties of many-body systems as external parameters, such as temperature or pressure, are varied \cite{A1}. It is now well understood, however, that the essential mechanisms underlying phase transitions are not restricted to thermal equilibrium. Analogous phenomena can emerge far from equilibrium, where singular behavior arises not in thermodynamic potentials, but in the dynamics of the system itself \cite{A2,A3,A4,A5,A6,A7,A8,A9,A10,A11,A12}.

In many cases, equilibrium phase transitions are accompanied by dynamical counterparts, especially when a control parameter is varied across or quenched through a critical point. The temporal evolution of such systems may exhibit non-analytic behavior -- abrupt changes in time-dependent observables -- signaling the presence of a dynamical phase transition \cite{A2,A3}. These transitions often arise at finite times and reflect underlying criticality in the non-equilibrium dynamics.

In the context of stochastic systems~\cite{A13,A14,A15,A16}, a recent work~\cite{A17} identified a particularly striking class of dynamical phase transitions associated with the relaxation dynamics toward stationarity. There, critical behavior is linked to eigenvalue crossings in the spectrum of the Markov generator under detailed balance. Specifically, such crossings can induce non-analytic shifts in the dominant decay mode, resulting in singular dynamical behavior formally analogous to first-order equilibrium phase transitions~\cite{A17}. These transitions are marked by discontinuities in the direction of relaxation, reflecting a sudden reorganization of the system's slowest modes.

However, for classical generators that satisfy detailed balance, the spectrum remains real and analytic with respect to system parameters. This rules out the possibility of eigenvector coalescence and, consequently, the occurrence of exceptional points -- non-Hermitian degeneracies where eigenvalues and eigenvectors simultaneously merge \cite{A18}. As a result, only first-order transitions can arise in such systems, imposing intrinsic constraints on the types of dynamical criticality they can support. 
\textcolor{black}{By contrast, non-Hermitian degeneracies are generic in dissipative quantum systems and can often be engineered or controlled with high precision~\cite{open}. Recent studies have demonstrated that complex second eigenvalues of Lindbladian generators can be exploited to accelerate relaxation~\cite{PRA2022,PRL133_140404,arXiv2507,arXiv2502,PRL131}, suggesting a connection between complex spectra and performance optimization in quantum dynamics. Such a speedup is closely related to, and can facilitate, the occurrence of the Mpemba effect (see e.g. \cite{
PRA2022,PRL133_140404, arXiv2502,Mpemba1,Mpemba2,Mpemba3} and reference therein.}
\textcolor{black}{Related phenomena have also been investigated in classical non-equilibrium settings where coupling to multiple thermal baths induces complex spectral behavior. In particular, Refs.~\cite{arXiv2502, PRL131} provide explicit examples where the relaxation spectrum acquires complex components and the role of complex second-eigenvalues is analyzed in detail.}

In this work, we apply the framework of dynamical phase transitions via eigenvalue crossing and coalescence to classical Markovian dynamics emerging from quantum systems undergoing repeated dephasing \cite{A19,A20,A21,A22,A23,A24,A25,A26,A27,A28,A29,A30,A31,A31b,A32,A33,A34,A34b,A35,A36}. Here, coherent unitary evolution is periodically interrupted by decoherence events \cite{A19,A20,A29}, effectively mapping the quantum dynamics onto a classical discrete-time Markov process. Unlike equilibrium systems \cite{A17}, these dynamics typically break detailed balance, resulting in a richer spectral structure.
\textcolor{black}{The main novelty of this work is the identification of both first- and second-order dynamical phase transitions within these dephased quantum systems. The latter are enabled by eigenvector coalescence at exceptional points, marking genuine second-order non-equilibrium transitions -- a feature absent in systems obeying detailed balance. These phenomena are illustrated in two paradigmatic models: a gauge-influenced quantum walk on a ring and a quantum walk with internal degrees of freedom on a finite line.}
\textcolor{black}{Importantly, the underlying models are experimentally realizable in photonic quantum walks, trapped ions, and ultracold atomic systems. These platforms offer a suite of laboratory tools for observing the predicted spectral transitions and relaxation phenomena, potentially opening new avenues for quantum sensing, error mitigation, and non-equilibrium control.}

\section{Phase Transitions via Eigenvalue Crossings in Classicalized Quantum Walks}
 
We consider a general linear quantum network consisting of $N$ nodes $|n\rangle$ ($n = 1, 2, \ldots, N$) and a single quantum particle undergoing continuous-time quantum walk dynamics via coherent hopping between the nodes. The evolution is governed by a single-particle Hamiltonian of the form  
\begin{equation}
H = \sum_{n,m} H_{n,m} |n\rangle\langle m|,
\end{equation}
where $H_{n,m} = H_{m,n}^*$. The unitary evolution of the system over a time interval $\tau$ is described by  
\begin{equation}
U = e^{-i H \tau} = \sum_{n,m} U_{n,m} |n\rangle\langle m|.
\end{equation}
The unitary matrix $U$ is assumed to depend on a controllable parameter $\beta$, which could be a tuning parameter entering in the Hamiltonian $H$ or just the coherent evolution time interval ($\beta= \tau$). After each time interval $\tau$, with probability $q$, a dephasing event occurs \cite{A19,A20,A29,A31b}. The system is described by a density matrix
\begin{equation}
\rho = \sum_{n,m} \rho_{n,m} |n\rangle\langle m|,
\end{equation}
which evolves in discrete time steps according to \cite{A29}
\begin{equation}
\rho^{(t+\tau)} = (1-q)\, U \rho^{(t)} U^\dagger + q \sum_n K_n U \rho^{(t)} U^\dagger K_n^\dagger,
\end{equation}
where the Kraus operators $K_n = |n\rangle\langle n|$ represent pure dephasing.
In this work, we mainly focus on the case $q = 1$, corresponding to dephasing at every step. In this limit, the dynamics become effectively classical, and the evolution reduces to a discrete-time Markov process for the diagonal elements of the density matrix. \textcolor{black}{The onset of phase transitions in the full quantum regime, i.e. for $q<1$, is discussed in the Appendix A. In the classical limit $q=1$, the Markov master equation reads} 
\begin{equation}
P_n^{(t+\tau)} = \sum_m Q_{n,m} P_m^{(t)},
\end{equation}
where $P_n^{(t)} = \rho^{(t)}_{n,n}$ denotes the probability of finding the particle at node $|n\rangle$ at time $t$, and $Q$ is the $N \times N$ Markov transition matrix with elements
\begin{equation}
Q_{n,m} = |U_{n,m}|^2.
\end{equation}
Indicating by $P^{(t)}=(P_1^{(t)}, P_2^{(t)},..., P_N^{(t)})^T$ the vector of probabilities, Eq.(5) can be written in a compact form as $P^{(t+\tau)}=QP^{(t)}$.
The off-diagonal element $Q_{n,m}$ ($n \neq m$) describes the probability that in one step the particle incoherently hops from node $m$ to node $n$ of the network, while $Q_{n,n}$ is the non-escaping probability at node $n$. For probability conservation, one clearly has $\sum_n Q_{n,m}=1$. 
Since $U$ is a unitary matrix, from Eq.(6) one also has $\sum_m Q_{n,m}=1$, i.e. $Q$ is a doubly stochastic matrix \cite{doubly}.  This implies that the stationary probability distribution $\pi$, i.e. such that $\pi=Q \pi$, is the uniform distribution $\pi=(1/N, 1/N,..., 1/N)^T$, independent of the system parameter $\beta$. If the stochastic matrix $Q$  is positive or regular, or for any irreducible aperiodic stochastic matrix \cite{doubly}, the stationary probability distribution is also unique.
The classical Markov process satisfies detailed balance provided that 
\begin{equation}
Q_{n,l} \pi_l =Q_{l,n} \pi_n.
\end{equation}
Taking into account that the stationary distribution $\pi_n$ does not depend on node $n$, detailed balance thus implies the symmetry constraint $Q_{n,l}=Q_{l,n}$. The detailed balance condition is satisfied whenever the coherent Hamiltonian $H$ has time reversal symmetry, i.e. $H_{n,m}=H_{m,n}$ is real, while it is generally not satisfied when $H$ does not have time reversal symmetry, for example in chiral networks when gauge fields are applied to the hopping rates among nodes of the network \cite{chiral}. It should be mentioned that, even when $H$ displays time reversal symmetry and thus the detailed balance condition is satisfied, the Markov matrix $Q$ is rather generally not embeddable \cite{A37}, i.e. it cannot always be written as the exponential of a rate matrix $G$, $Q=\exp(G)$, with $G$ in the Arrhenius form. This means that, even though the quantum walk on the network is fully classicalized by periodic dephasing and the resulting Markov stochastic process satisfies detailed balance, it cannot be rather generally reduced to a continuous-time stochastic process with a rate matrix $G$ in the Arrhenius form considered in previous works \cite{A17}.
\begin{figure}
\includegraphics[width=18 cm]{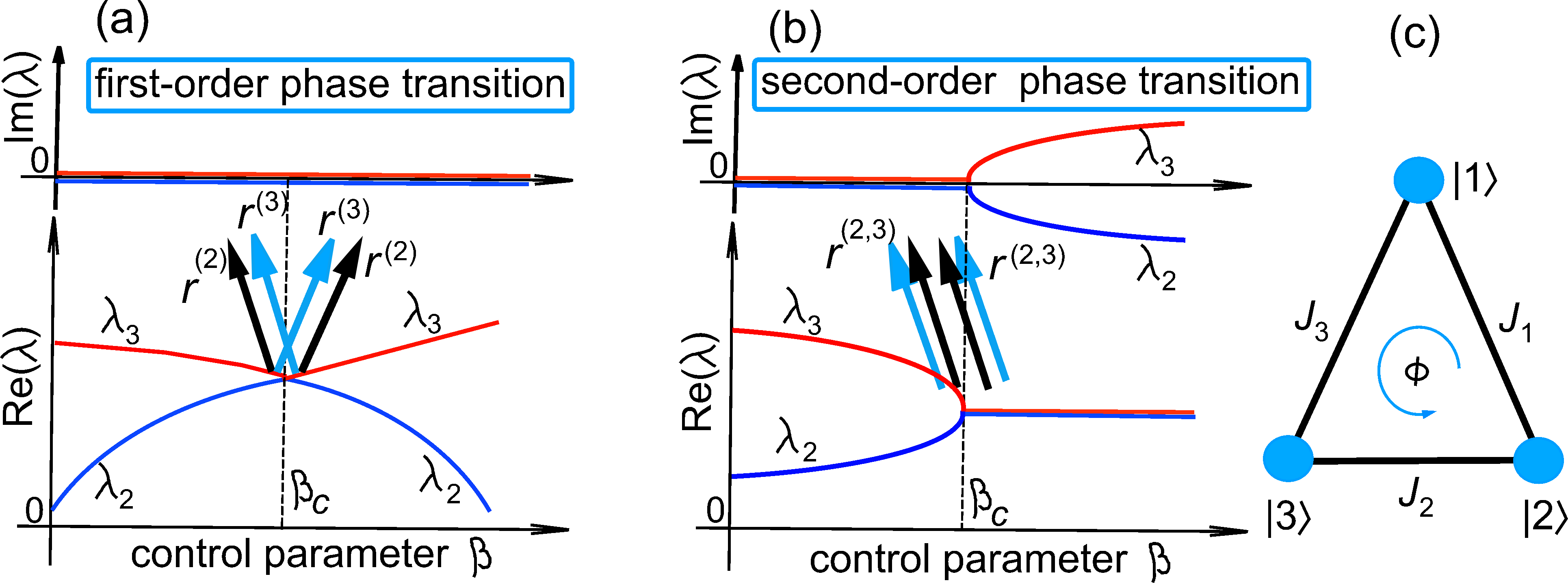}
\caption{Dynamical phase transitions in the decohered quantum walk on a network. The panels (a) and (b) show typical behaviors of the real (lower panels) and imaginary parts (upper panels) of the Floquet exponents $\lambda_2$ and $\lambda_3$ of the first two dominant decay modes versus the control parameter $\beta$. At the critical value $\beta=\beta_c$ the two eigenvalues coalesce, i.e. $\lambda_2=\lambda_3$.  Under time reversal symmetry, detailed balance is satisfied and the eigenvalue crossing corresponds to a jump of the right eigenvector $r^{(2)}$ (first-order phase transition),  as shown in (a). Detailed balance can be violated when time reversal symmetry is broken. In this case eigenvalue crossing can correspond to a coalescence of both eigenvalues and eigenvectors, i.e. to an exceptional point, as shown in (b). In this case the right eigenvector $r^{(2)}$, i.e. direction to equilibration, is continuous as the critical point $\beta=\beta_c$ is crossed, leading to a dynamical analogue of a second-order phase transition.
 (c) Minimal network for the observation of first- and second-order dynamical phase transitions. The network comprises three nodes $| 1 \rangle$, $|2  \rangle$ and $| 3 \rangle$, with coherent hopping rates $J_1$, $J_2$ and $J_3$. A gauge field (magnetic flux) $ \phi$ can be applied to the closed triangular path. Time reversal symmetry, and thus detailed balance, is violated for $ \phi \neq 0, \pi$.}
\end{figure}
To highlight the onset of dynamical phase transitions in the relaxation toward the equilibrium distribution \cite{A17}, let us indicate by $\mu_n=\exp(- \lambda_n)$ the eigenvalues and by $\lambda_n$ the Floquet exponents of the Markov transition matrix $Q$, ordered such that  $\lambda_1=0< {\rm Re(\lambda_2}) \leq {\rm Re(\lambda_3}) \leq ... \leq  {\rm Re(\lambda_N})$, and by $r{^{(n)}}$, $l{^{(n)}}$ the corresponding right and left eigenvectors, i.e. $Q r{^{(n)}}=\mu_n r{^{(n)}}$ and $Q^{\dag} l{^{(n)}}= \mu_n^*  l{^{(n)}}$. The eigenvalue $\mu_1=1$ corresponds to the stationary state $\pi$, with right and left eigenvectors $r{^{(1)}}= \pi$, and $l{^{(1)}}=1$, whereas ${\rm Re}(\lambda_n)>0$ is the decay rate of the $n-$th decaying mode ($n \geq 2$). Note that the imaginary part of the Floquet exponents $\lambda_n$ is defined up to integer multiples of $ 2 \pi$. Assuming the orthonormality condition $ \langle l^{(l)}| r^{(s)}\rangle= \delta_{l,s}$, any initial non-equilibrium probability distribution $P^{(t=0)}$ relaxes toward the unique stationary uniform distribution $\pi$ according to
\begin{equation}
P^{(t)}=\pi +\sum_{s=2}^N C_s r^{(s)} \exp(-\lambda_s t / \tau)
\end{equation}
 where $t=0, \tau, 2 \tau, 3 \tau,...$ and where $C_s=\langle l^{(s)}| P^{(0)} \rangle$ are the spectral amplitudes. The second eigenvalue sets an exponential timescale of the relaxation $ \sim 1 / {\rm Re}(\lambda_2)$. If ${\rm Re}(\lambda_2)$ is strictly smaller than ${\rm Re}(\lambda_3)$ and $C_2 \neq 0$, the long-time stage
of the relaxation process is in the direction of $r^{(2)}$, and it changes continuously with the control parameter $\beta$. However, if there is an eigenvalue crossing and at some critical value $\beta=\beta_c$ the decay rates of second and third mode become equals, i.e. ${\rm Re}(\lambda_3)={\rm Re}(\lambda_2)$ at $\beta=\beta_c$, a dynamical phase transition is observed as $\beta$ is varied across the critical value. To characterize the nature of the phase transitions, we need to distinguish between two cases. The first case corresponds to the scenario studied in Ref. \cite{A17}, where the Markov matrix $Q$ satisfies detailed balance, i.e., $H$ exhibits time-reversal symmetry. In this case, all eigenvalues $\mu_n$ are real, either positive or negative, implying that the imaginary part of $2 \lambda_n$ is always vanishing (mod. $ 2 \pi$) and can be thus disregarded. The critical point  $\beta=\beta_c$ corresponds to the coalescence of second and third eigenvalues, namely $ {\rm Re} (\lambda_2)={\rm Re}( \lambda_3)$, with distinct right eigenvectors $r^{(2)} \neq r^{(3)}$ [Fig.1(a)].  Crossing the critical point $\beta=\beta_c$ leads to a sharp {\em discontinuity} in the final direction of the approach to equilibrium, from $r^{(2)}$ to $r^{(3)}$, and this behavior can be interpreted as a first-order dynamical phase transition \cite{A17}. However, when $H$ breaks time-reversal symmetry and thus detailed balance, the eigenvalues $\mu_n$ can become complex and the crossing point $\beta=\beta_c$ can correspond to an exceptional point \cite{A18}, with the simultaneous coalescence of both eigenvalues $\lambda_2=\lambda_3$ and eigenvectors $r^{(2)}=r^{(3)}$. In this case, as the critical point $\beta=\beta_c$ is crossed, the pair of eigenvalues $\lambda_{2,3}$ become complex conjugate one another, as schematically illustrated in Fig.1(b). Correspondingly, the long-term relaxation dynamics does not exhibit a discontinuous jump in the direction of relaxation, since $r^{(2)}=r^{(3)}$ at the transition point and the direction of relaxation thus changes continuously with control parameter $\beta$. Specifically, one observes a continuous transition from monotonous to oscillatory behavior as the system approaches equilibration. This behavior can thus be interpreted as a second-order dynamical phase transition.\\
 To illustrate the onset of first- and second-order dynamical phase transitions in the  relaxation dynamics toward equilibrium, let us consider the minimal network composed by three nodes $|1 \rangle$, $|2 \rangle$ and $|3 \rangle$, as schematically shown in Fig.1(c). We indicate by $J_1$, $J_2$ and $J_3$ the coherent hopping amplitudes among the three nodes, and assume a gauge field (magnetic flux) $\phi$ along the triangular closed path of the network. The Hamiltonian describing coherent evolution reads
 \begin{equation}
 H= J_1 |1 \rangle \langle 2|+J_2 |2 \rangle \langle 3 | + J_3 \exp(i \phi) |1 \rangle \langle 3| +{\rm H.c.}
 \end{equation}
 \begin{figure}
\includegraphics[width=19 cm]{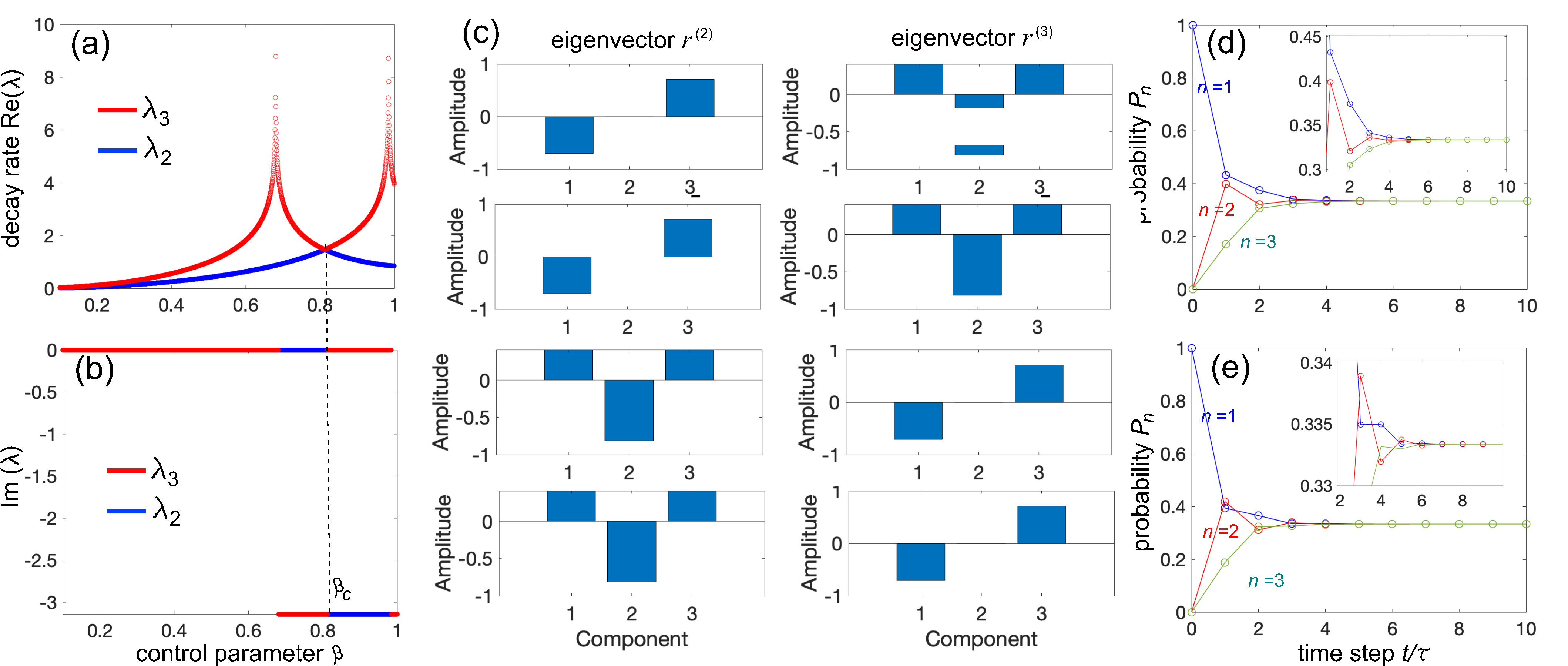}
\caption{First-order dynamical phase transitions in the minimal network of Fig.1(c) with time-reversal symmetry ($\phi=0$) for parameter values $J_1=J_2=1$ and $J_3=0.5$. The control parameter  is $\beta=J_1 \tau= \tau$.  (a) and (b) show the behaviors of the real [panel (a)] and imaginary parts [panel (b)]  of the Floquet exponents $\lambda_2$ and $\lambda_3$ of the first two dominant decay modes versus the control parameter $\beta$. At the critical value $\beta=\beta_c \simeq 0.8127$ the two eigenvalues coalesce, i.e. $\lambda_2=\lambda_3$, while the respective eigenvectors are distinct. Crossing the critical point yields a discontinuity in the direction to equilibration. \textcolor{black}{(c) Behavior of the right eigenvectors $r^{(2)}$ and $r^{(3)}$ for a few increasing values of $\beta$, showing the flipping of the eigenvectors as $\beta$ crosses the critical value. From top to bottom: $\beta=0.7$, $\beta=0.8$, $\beta=0.83$, and $\beta=1$}. (d,e) Relaxation dynamics, showing the behaviors of the node occupation probabilities $P_n$ at successive time steps, for the initial condition $P_n^{(t=0)}=\delta_{n,1}$, corresponding to initial particle occupying site $|1 \rangle$. In (c) $\beta=0.79< \beta_c$, whereas in (d) $\beta=0.83> \beta_c$. The insets in (d) and (e) display an enlargement of the long-time relaxation dynamics.}
\end{figure}
 As a control parameter $\beta$, we take the coherent time interval $\tau$ normalized to the hopping rate $J_1$, i.e. $\beta=J_1 \tau$. The Hamiltonian $H$ does not possess time-reversal symmetry whenever $ \phi \neq 0, \pi$. This system could be realized, for example, in photonic platforms using three coupled optical waveguides with periodic dephasing segments along the propagation axis \cite
{A32}. Figures 2(a) and (b) show typical behaviors of the real and imaginary parts of the two eigenvalues $\lambda_2$ and $\lambda_3$ versus $\beta$ for parameter values  $J_1=J_2=1$, $J_3=0.5$ and $\phi=0$, corresponding to a symmetric Markov matrix $Q$ satisfying detailed balance. As one can see, there is an eigenvalue crossing at $\beta=\beta_c \simeq 0.8127$, and the two eigenvectors $r^{(2)}$ and $r^{(3)}$ are distinct at $\beta=\beta_c$, namely one has $r^{(2)}=(1/ \sqrt{2},0,-1/ \sqrt{2})^T$ and $r^{(3)}=(1/\sqrt{6},-2 / \sqrt{6}, 1/ \sqrt{6})^T$. \textcolor{black} {Figure 2(c) shows the evolution of the right eigenvectors $r^{(2)}$ and $r^{(3)}$ as the control parameter $\beta$ is spanned across the critical value $\beta_c$}. Since $Q$ is a symmetric real matrix, the left and right eigenvectors do coincide. Crossing the critical point $\beta=\beta_c$ thus leads to a {\em discontinuity} in the final direction of the approach to equilibrium \cite{A17}. In particular, for $\beta< \beta_c$ the dominant decaying mode is $r^{(2)}$ with a vanishing amplitude of its second element as $\beta \rightarrow \beta_c^-$, indicating that the occupation probability of node 2 , i.e. $P_2^{(t)}$,in the long-time limit should relax to its equilibrium value $\pi_2=1/3$ faster than the occupation probabilities of the other two sites. Conversely, for $\beta> \beta_c$ the dominant decay mode is $r^{(3)}$, which has a non-vanishing amplitude at all three sites: in this case the long-time relaxation dynamics should display a similar decay behavior for the three sites. 
This scenario is illustrated by comparing the relaxation dynamics from an initial non-equilibrium state, for example, the state  
$ P^{(t=0)} = (1, 0, 0)^T$,
which corresponds to an initial excitation of node 1. The dynamics are examined for two values, $\beta = \beta_c^-$ and $\beta = \beta_c^+$, as shown in Figs.~2(d) and 2(e), respectively.
In Fig.~2(d), corresponding to $\beta = 0.79$, which is slightly below the critical value $\beta_c$, the long-time behavior -- highlighted in the inset -- reveals a noticeably faster relaxation of the occupation probability for site 2 compared to sites 3 and 4. In contrast, this distinction is absent in Fig.~2(e), where $\beta = 0.82 > \beta_c$.

To observe a second-order dynamical phase transition, we introduce a non-vanishing gauge phase $\phi = \pi/3$. Figures~3(a) and 3(b) show the behavior of the real and imaginary parts of the two Floquet exponents $\lambda_2$ and $\lambda_3$ as functions of $\beta$, for the same parameters as in Fig.~2, except that $\phi = \pi/3$. The plots clearly indicate the emergence of an exceptional point at the critical value $\beta_c \simeq 0.7412$, where not only do the eigenvalues $\lambda_2$ and $\lambda_3$ coalesce, but their corresponding right eigenvectors also merge [Fig.3(c)]. Specifically, at $\beta = \beta_c$, one finds
$ r^{(2)} = r^{(3)} = (0.7887,\,-0.5773,\,-0.2113)^T.$
In this case, the relaxation dynamics toward equilibrium undergo a smooth yet qualitatively distinct transition as $\beta$ crosses $\beta_c$: from monotonous relaxation below the critical point to oscillatory relaxation above it, as illustrated in Figs.~3(d) and 3(e). \textcolor{black}{The distinction between first- and second-order transitions is clearly reflected in the relaxation dynamics shown in panels (d) and (e) of Figs. 2 and 3. In the first-order case, the relaxation trajectory exhibits a sharp change at \( \beta_c \), consistent with a discontinuous switching of the dominant decay mode. This is evidenced by a faster damping of the population at site \( n = 2 \), compared to the other two sites, for \( \beta < \beta_c \). In contrast, in the second-order case, the system approaches a non-diagonalizable regime near the exceptional point, leading to slow, non-exponential relaxation dynamics. The behavior on either side of the exceptional point is also qualitatively different: for $\beta < \beta_c$, the system exhibits nearly monotonic relaxation, while for $ \beta > \beta_c$, the relaxation dynamics become oscillatory due to the emergence of complex-conjugate eigenvalues.}
 \begin{figure}
\includegraphics[width=19 cm]{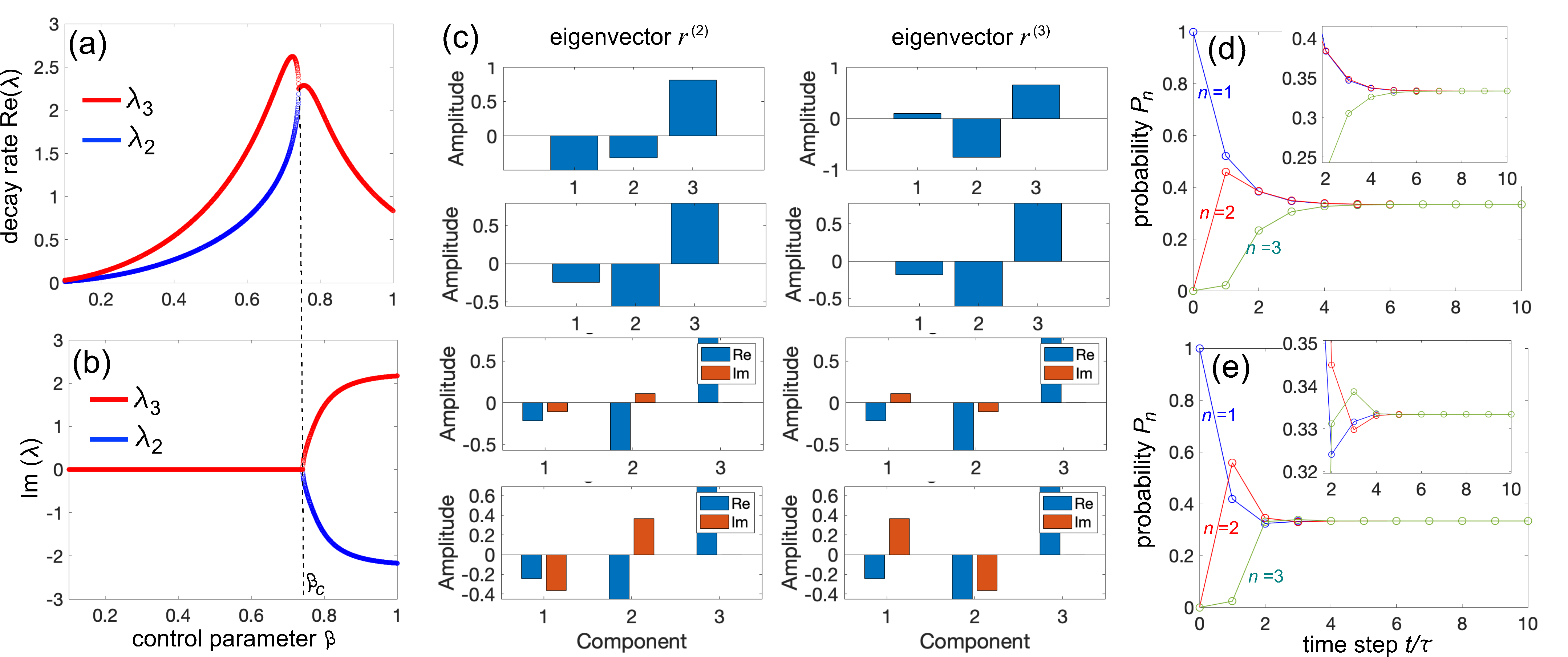}
\caption{Same as Fig.2, but for a non-vanishing gauge field $\phi= \pi/3$. Time-reversal symmetry breaking yields a Markov matrix which violates detailed balance. An exceptional point is observed at the critical value $\beta=\beta_c=0.7412$. Crossing the critical point corresponds to a second-order (smooth) phase transition in the approach to equilibrium. \textcolor{black}{Panel (c) shows the behavior of the right eigenvectors $r^{(2)}$ and $r^{(3)}$ of $Q$ for a few increasing values of $\beta$ crossing the critical point (from top to bottom: $\beta=0.6$, $\beta=0.74$, $\beta=0.76$, and $\beta=1$). Note the  smooth change of the dominant decay eigenmode and eigenvector coalescence as the critical point is crossed.} Panels (d) and (e) show the relaxation dynamics for $\beta=0.7 <\beta_c$ and $\beta=0.8 >\beta_c$, respectively.}
\end{figure}
\section{Emergence of Second-Order Dynamical Phase Transitions in Quantum Walks with Internal States}
\textcolor{black}{ In this section, we study the emergence of second-order dynamical phase transitions in the relaxation dynamics of discrete-time quantum walks involving a quantum particle with an internal state (such as spin or polarization)~\cite{A19,caz1,caz2,caz3}. Importantly, it is shown that the breaking of detailed balance -- and thus the occurrence of second-order transitions characterized by exceptional points -- can arise solely due to the presence of internal degrees of freedom, in the absence of external gauge fields.} These types of discrete-time quantum walks are experimentally accessible on a range of platforms, including trapped ions~\cite{trap1,trap2,trap3}, cold atoms~\cite{A31b,cold1}, and photons~\cite{A29,A30,phot1,phot2,phot3,photo3b,photo3c,phot4,phot5}. Specifically, I focus on discrete-time quantum walks on a finite one-dimensional lattice with reflecting boundary conditions~\cite{A38,A39,A39b},  where symmetry breaking, \( Q_{n,l} \neq Q_{l,n} \), occurs without the need for gauge phases. The quantum walker possesses two internal states, such as spin or polarization, denoted by \( |H \rangle \) and \(  |V \rangle \), and moves along discrete positions \( n = 1, 2, \dots, L \) on a line. The coherent unitary evolution in one step, from time $t$ to $(t+ \tau)$, is defined by the unitary operator
\begin{equation}
U=S_{V} C S_{H}
\end{equation}
where 
\[ S_{H} = \sum_n \left( |n-1 \rangle \langle n| \otimes | H \rangle \langle H |+  |n \rangle \langle n| \otimes | V \rangle \langle  V | \right) \]
shifts the walker with internal state $|  H \rangle$ to the left, i.e. from site $| n \rangle$ to site $|n-1 \rangle$,
\[ S_{V} = \sum_n \left( |n \rangle \langle n| \otimes | H \rangle \langle H |+  |n+1 \rangle \langle n| \otimes | V \rangle \langle V | \right) \]
shifts the walker with internal state $| V \rangle$ to the right, i.e. from site $| n \rangle$ to site $|n+1 \rangle$, and
\begin{eqnarray}
C & = &  
\begin{pmatrix}
\cos \beta_n & - \sin \beta_n \\
 \sin \beta_n & \cos \beta_n
\end{pmatrix}= \sum_n \cos \beta_n  | n \rangle \langle n| \otimes   (|H \rangle \langle H | +  |V \rangle \langle V|) \nonumber \\
& + & \sum_{n }\sin {\beta_n} | n \rangle \langle n| \otimes ( |V \rangle \langle H| -  |H \rangle \langle V|) 
\end{eqnarray}
 is the coin operator at spatial position $n$ that rotates the internal state by an angle $\beta_n$.
 After each time step, decoherence with probability $q$ is applied, so that the discrete-time evolution of the density operator is governed by the map
 \begin{equation}
 \rho^{(t+ \tau)}= (1-q) U \rho^{(t)}U^{\dag} +q \sum_{n, \sigma=H,V} K_{n,\sigma} U \rho^{(t)}U^{\dag} K_{n, \sigma}^{\dag}
 \end{equation}
 where $K_{n, \sigma}= |n \rangle  \langle n | \otimes | \sigma \rangle \langle \sigma|$ ($\sigma=H$ or $V$) are the Kraus operators for pure dephasing.  For $q = 1$, the dynamics becomes effectively classical and the evolution reduces to a Markov process for the diagonal elements of the density matrix, i.e. for the probabilities $X_n^{(t)}$ and \( Y_n^{(t)}  \) to find the walker at position \( n \) on the line with internal state \( |H \rangle \) and \( |V\rangle \),  respectively. The Markov process reads explicitly ~\cite{A35,A36,figo}
\begin{eqnarray}
 X_n^{(t+\tau)} & = & \cos^2 \beta_n X_{n+1}^{(t)}+ \sin^2 \beta_n Y_{n}^{(t)}  \\
 Y_n^{(t+\tau)} & = & \sin^2 \beta_{n-1} X_{n}^{(t)}+ \cos^2 \beta_{n-1} Y_{n-1}^{(t)}.
\end{eqnarray}
 To realize reflection at the edge spatial positions $n=1,L$, we assume $\beta_n= \pi/2$ for $n=0,L$ \cite{A35,A38,A39}, while a uniform rotation angle $\beta_n=\beta$ is assumed for $n=1,2,...,L-1$. After letting $P^{(t)}=(X_1^{(t)},X_2^{(t)},...,X_L^{(t)},Y_1^{(t)},..., Y_L^{(t)} )^T$, Eqs.(13) and (14) can be written as $P^{(t+\tau)}=Q P^{(t)}$, where $Q$ is the $(N \times N)$ Markov transition matrix and $N=2L$.
 It can be readily shown that $Q$ is a a doubly stochastic matrix, i.e. $\sum_n Q_{n,m}=\sum_m Q_{n,m}=1$, however it is not symmetric, $Q_{n,m} \neq Q_{m,n}$. Also, the $N=2L$ eigenvalues $\mu=\exp(-\lambda)$ of $Q$ with Floquet exponents $\lambda$ appear in pairs, namely if $\mu$ is an eigenvalue of $Q$ with right eigenvector $r=(x,y)^T$, then $\mu'=-\mu$ (i.e. $\lambda'=\lambda+ i \pi$) is also an eigenvalue of $Q$ with right eigenvector $r'=(x',y')^T$, where $x'_l=(-1)^lx_l$, $y'_l=(-1)^{l+1}y_l$ ($l=1,2,..,L$).  In particular, along with the eigenvalue $\mu_1=1$, corresponding to the non-decaying stationary probability distribution $\pi=(x,y)^T=(1/N,1/N,...,1/N)^T$, there is a companion non-decaying eigenstate with eigenvalue $\mu_2=-1$, so that ${\rm Re}(\lambda_1)={\rm Re}(\lambda_2)=0$. This implies that, starting from an arbitrary initial probability distribution, the system does not relax to a stationary (equilibrium) state. Instead, the distribution $P^{(t)}$ exhibits persistent oscillatory behavior due to the interference of two non-decaying modes. This behavior corresponds to a period-doubling symmetry in the internal dynamics, leading to persistent alternation between internal states rather than convergence.
 However, if one considers the marginal probability distribution $p_l^{(t)} = X_l^{(t)} + Y_l^{(t)}$, i.e., the probability of finding the particle at position $l$ at time $t$, regardless of its internal state, then $p^{(t)}$ relaxes to the uniform equilibrium distribution $p^{(\infty)} = (1/L, 1/L, \dots, 1/L)^T$. This occurs because the mode with eigenvalue $\mu_2 = -1$ does not contribute to the relaxation dynamics of $p_l^{(t)}$.
To unveil the occurrence of a second-order phase transition, we assume the coin angle $\beta$ as a control parameter, and calculate the behaviors of the real and imaginary parts of the decaying modes associated to the Floquet exponents $\lambda_l$, $l=3,4,..,2L$, versus $\beta$. A typical eigenvalue behavior for a minimal quantum walk comprising $L=3$ positional sites on the line is shown in Figs.4(a) and (b), clearly indicating the appearance of an exceptional point at $\beta=\beta_c \simeq 0.4771 \times \pi/2$. A similar behavior is observed for a larger number $L$ of positional sites, with the critical parameter $\beta_c$ decreasing as $L$ increases, as illustrated in the inset of Fig.4(b). The relaxation dynamics toward the equilibrium uniform distribution $p^{(\infty)}=(1/3,1/3,1/3)^T$ exhibits a smooth transition as $\beta$ crosses the critical point. Specifically, the relaxation pattern changes  from oscillatory [Fig.4(c), $ \beta< \beta_c$] to monotonic [Fig.4(d), $ \beta> \beta_c$] behavior.\\

 \begin{figure}
\includegraphics[width=14 cm]{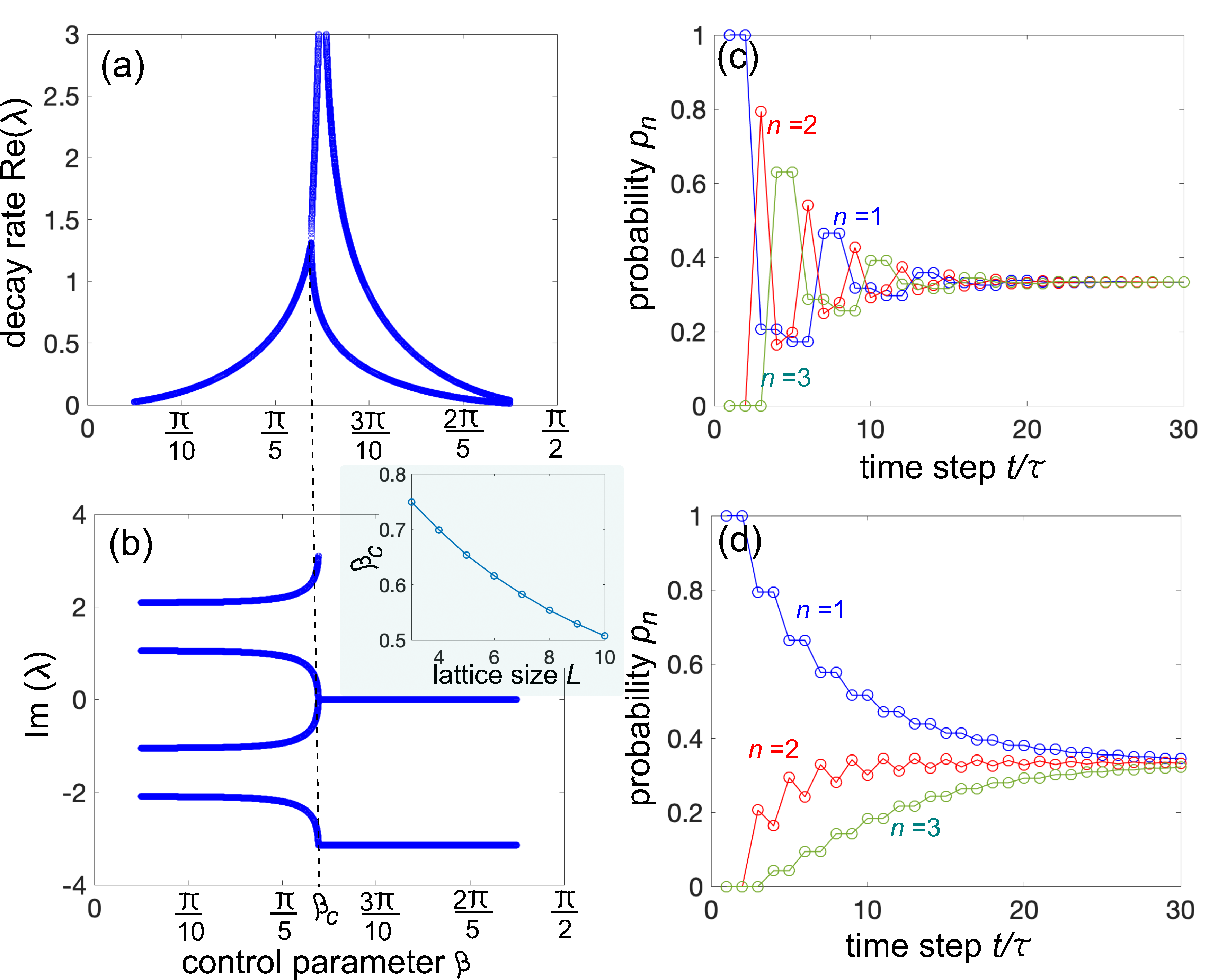}
\caption{Second-order dynamical phase transition in the discrete-time quantum walk on a line (size $L=3$) with reflective boundaries under dephasing. The walker can be found at the three spatial positions $n=1,2$ and 3 with either internal state $|H \rangle$ or $|V \rangle$. (a,b) Behavior of the Floquet exponents $\lambda_l$ ($l=3,4,5,6$) of the four decay modes (real and imaginary parts) versus coin angle $\beta$. An exceptional point occurs at the critical value $\beta=\beta_c \simeq 0.4771 \times \pi/2$. The critical coin angle $\beta_c$ depends on the system size $L$, as shown in the inset of (b). (c,d) Relaxation dynamics toward equilibrium distribution for a coin angle $\beta$ below [$\beta=0.3 \times \pi/2$, panel (c)] and above [$\beta=0.7 \times \pi/2$, panel (d)]] the critical value $\beta_c$. The initial probability distribution is $X_1=1$ and $X_l=Y_l=0$  otherwise, corresponding to the walker placed at site $n=1$ with internal state $| H \rangle$. The curves in the plots show the site occupation probabilities $p_n=X_n+Y_n$ of the walker at the three sites $n=1,2,3$ versus time step, regardless of its internal state.}
\end{figure}
  
\section{Conclusion}

In this work, we have uncovered a new class of dynamical phase transitions arising from the classicalization of quantum system dynamics through repeated dephasing, where the breakdown of detailed balance enables second-order transitions. I analyzed two paradigmatic systems -- a gauge-influenced quantum walk on a ring and an internally structured walk on a finite line -- both exhibiting dynamical phase transitions via broken detailed balance under dephasing. These results deepen our understanding of critical relaxation behavior in non-equilibrium quantum systems and open new avenues for experimental investigation across various quantum platforms. \textcolor{black}{A relevant aspect of this work is to suggest the occurrence of both first- and second-order phase transitions in technologically accessible platforms, such as photonic quantum walks, which could provide a versatile set of laboratory tools to directly observe the predicted relaxation phenomena.}  
Further, our findings bear direct relevance to advanced quantum technologies. Quantum walks underpin key applications in quantum simulation, information processing, and transport. Insights into how decoherence-induced phase transitions influence relaxation dynamics can inform the design of more robust quantum devices that either exploit or mitigate decoherence effects. In particular, controlling and probing such dynamical phase transitions may enable novel approaches to error correction, quantum sensing, and the engineering of non-equilibrium quantum states with enhanced coherence. Finally, the experimental feasibility of these models on platforms like trapped ions, photonic lattices, and ultracold atoms paves the way for tangible exploration of these phenomena.

Looking forward, extending this framework to interacting multi-particle quantum walks, higher-dimensional systems, and more complex noise models promises to enrich our understanding of critical phenomena in far-from-equilibrium quantum systems. \textcolor{black}{Moreover, the acceleration of relaxation dynamics linked to complex eigenvalues, closely related to effects such as the Mpemba effect~\cite{PRA2022,PRL133_140404, arXiv2502,Mpemba1,Mpemba2,Mpemba3}, suggests further intriguing directions for research.} Overall, this work advances the theoretical foundation of dynamical criticality in open quantum systems and lays important groundwork for leveraging these effects in next-generation quantum technologies.


\medskip
\textbf{Conflict of Interest} \par
The author declares no conflict of interest.

\medskip
\textbf{Data Availability Statement}
The data that support the findings of this study are available from the corresponding author upon reasonable request. \par

\medskip
\textbf{Acknowledgements} \par 
The author acknowledges the Spanish State Research Agency, through
the Severo Ochoa and Maria de Maeztu Program for Centers 
and Units of Excellence in R\&D (Grant No. MDM-2017-
0711).

\medskip

%

\newpage
\begin{thebibliography}{99}

\bibitem{A1}
S. Sachdev, \textit {Quantum Phase Transitions} (Cambridge University Press, Cambridge, UK, 2011).
\bibitem{A2}
M. Heyl. \textit {Dynamical quantum phase transitions: a review}, Rep.Prog. Phys.
{\bf 81}, 054001 (2018).
\bibitem{A3}
 M. Heyl, \textit  {Dynamical quantum phase transitions: a survey}, EPL
{\bf 125}, 26001 (2019).
\bibitem{A4}
M. Heyl, A. Polkovnikov, and S. Kehrein, {\it Dynamical Quantum
Phase Transitions in the Transverse-Field Ising Model}, Phys.
Rev. Lett. {\bf 110}, 135704 (2013).
\bibitem{A5}
M. Heyl, {\it Dynamical Quantum Phase Transitions in Systems with
Broken-Symmetry Phases}, Phys. Rev. Lett. {\bf 113}, 205701 (2014).
\bibitem{A6}
J. Eisert, M. Friesdorf, and C. Gogolin, {\it Quantum many-body systems out of equilibrium,}, Nat. Phys. {\bf 11}, 124 (2015).
\bibitem{A7}
P. Jurcevic, H. Shen, P. Hauke, C. Maier, T. Brydges, C. Hempel,
B. P. Lanyon, M. Heyl, R. Blatt, and C. F. Roos, {\it Direct Observation of Dynamical Quantum Phase Transitions in an Interacting
Many-Body System}, Phys. Rev. Lett. {\bf 119}, 080501 (2017).
\bibitem{A8}
J. Lang, B. Frank, and J. C. Halimeh, {\it Dynamical Quantum
Phase Transitions: A Geometric Picture}, Phys. Rev. Lett. {\bf 121},
130603 (2018).
\bibitem{A9}
K. Wang, X. Qiu, L. Xiao, X. Zhan, Z. Bian, W. Yi, and P. Xue,
\textit {Simulating Dynamic Quantum Phase Transitions in Photonic Quantum Walks}, Phys. Rev. Lett. {\bf 122}, 020501 (2019).
\bibitem{A10}
J. A Muniz, D. Barberena, R. J. Lewis-Swan, D. J. Young, J. R.
K. Cline, A. M. Rey, and J. K. Thompson, {\it Exploring dynamical
phase transitions with cold atoms in an optical cavity}, Nature
{\bf 580}, 602 (2020).
\bibitem{A11}
 S. De Nicola, A. A. Michilidis, and M. Serbyn, {\it Entanglement
View of Dynamical Quantum Phase Transitions}, Phys. Rev. Lett.
{\bf 126}, 040602 (2021).
\bibitem{A12}
J. Marino, M. Eckstein, M. S. Foster, and A. M. Rey, {\it Dynamical phase transitions in the collisionless pre-thermal states of
isolated quantum systems: theory and experiments}, Rep. Prog.
Phys. {\bf 85}, 116001 (2022).


\bibitem{A13}
J. Schnakenberg, {\it Network Theory of Microscopic and
Macroscopic Behavior of Master Equation Systems}, Rev.
Mod. Phys. {\bf 48}, 571 (1976).
\bibitem{A14}
R.A. Blythe,
{\it An introduction to phase transitions in stochastic dynamical systems}, J. Phys.: Conf. Ser. {\bf 40}, 1 (2006).
\bibitem{A15}
O. Raz, Y. Subasi, and C. Jarzynski, {\it  Mimicking Nonequilibrium Steady States with Time-Periodic Driving},
Phys. Rev. X {\bf 6}, 021022 (2016).
\bibitem{A16}
M.F. Weber and E. Frey, {\it Master equations and the theory of stochastic path integrals},
Rep. Prog. Phys. {\bf 80}, 046601 (2017).


\bibitem{A17}
G. Teza, R. Yaacoby, and O. Raz, {\it Eigenvalue Crossing as a Phase Transition in Relaxation
Dynamics}, Phys. Rev. Lett. {\bf 130}, 207103 (2023).

\bibitem{A18}
W.D. Heiss, {\it The physics of exceptional points}, J. Phys. A {\bf 45}, 444016 (2012).

\bibitem{open}
W. Chen, M. Abbasi, B. Ha, S. Erdamar, Y.N. Joglekar, and K.W. Murch, {\it Decoherence-Induced Exceptional Points in a Dissipative Superconducting Qubit},
Phys. Rev. Lett. {\bf 128}, 110402 (2022).

\textcolor{black}{
\bibitem{PRA2022}
S. Kochsiek, F. Carollo, and I. Lesanovsky,
\textit{Accelerating the approach of dissipative quantum spin systems towards stationarity through global spin rotations},
Phys. Rev. A {\bf 106}, 012207 (2022). 
\bibitem{PRL133_140404}
M. Moroder, O. Culhane, K. Zawadzki, and J. Goold,
\textit{Thermodynamics of the quantum Mpemba effect},
Phys. Rev. Lett. {\bf 133}, 140404 (2024).
\bibitem{arXiv2507}
N. Beato and G. Teza, {\it Relaxation control of open quantum systems}
arXiv:2507.15948 (2025). 
\bibitem{arXiv2502}
G. Teza, J. Bechhoefer, A. Lasanta, O. Raz, and M. Vucelja,
\textit{Speedups in nonequilibrium thermal relaxation: Mpemba and related effects},
arXiv:2502.01758 (2025). 
\bibitem{PRL131}
G. Teza, R. Yaacoby, and O. Raz,
\textit{Relaxation Shortcuts through Boundary Coupling},
Phys. Rev. Lett. {\bf 131}, 017101 (2023). 
\bibitem{Mpemba1}
F. Carollo, A. Lasanta, and I. Lesanovsky,
{\it Exponentially accelerated approach to stationarity in Markovian open quantum systems through the Mpemba effect},
Phys. Rev. Lett. {\bf 127}, 060401 (2021).
\bibitem{Mpemba2}
S. Longhi, {\it Mpemba effect and super-accelerated thermalization in the damped quantum harmonic oscillator},
Quantum {\bf 9}, 1677 (2025).
\bibitem{Mpemba3}
F. Ares, P. Calabrese, and S. Murciano,  {\it The quantum Mpemba effects}, Nature Phys. Rev. {\bf 7}, 451 (2025).
}



\bibitem{A19}
 V. Kendon, {\it Decoherence in quantum walks -- a review}, 
Mathematical Structures in Comp. Sci. {\bf 17}, 1169 (2007).
\bibitem{A20}
T.A. Brun, H.A. Carteret, and A. Ambainis,  
 {\it Quantum to Classical Transition for Random Walks},
Phys. Rev. Lett. {\bf 91}, 130602 (2003). 
\bibitem{A21}
D. Shapira, O. Biham, A.J. Bracken, and M. Hackett, {\it One-dimensional quantum walk with unitary noise}, Phys. Rev. A {\bf 68}, 062315 (2003).
\bibitem{A22}
V. Kendon and B.C. Sanders, {\it Complementarity and quantum walks}, Phys. Rev. A {\bf 71},  022307 (2005).
\bibitem{A23}
A. Romanelli, R. Siri, G.  Abal, A. Auyuanet, and R. Donangelo, {\it Decoherence in the quantum walk on the line}, Physica A {\bf 347}, 137 (2005).
\bibitem{A24}
 J. Kosik, V. Buzek, and M. Hillery, {\it Quantum walks with random phase shifts}, Phys. Rev. A {\bf 74},  022310 (2006).
 \bibitem{A25} 
M.B. Plenio and S.F. Huelga,
{\it Dephasing-assisted transport: quantum networks and biomolecules},
New J. Phys. {\bf 10}, 113019 (2008).
\bibitem{A26}
P. Rebentrost, M. Mohseni, I. Kassal, S. Lloyd, and A. Aspuru-Guzik,
{\it Environment-assisted quantum transport},
New J. Phys. {\bf 11},  033003 (2009).
 \bibitem{A27}
 M. Annabestani, S.J. Akhtarshenas, and M.R. Abolhassani,  {\it Decoherence in a one-dimensional quantum walk},
Phys. Rev. A {\bf 81}, 032321 (2010).
\bibitem{A28}
J.D. Whitfield, C.A. Rodriguez-Rosario, and A. Aspuru-Guzik,
{\it Quantum stochastic walks: A generalization of classical random walks and quantum walks},
Phys. Rev. A {\bf 81}, 022323 (2010).
\bibitem{A29}
M. A. Broome, A. Fedrizzi, B. P. Lanyon, I. Kassal, A. Aspuru-Guzik, and A. G. White, {\it Discrete Single-Photon Quantum Walks with Tunable Decoherence},
Phys. Rev. Lett. {\bf 104}, 153602 (2010).
\bibitem{A30}
A. Schreiber, K. N. Cassemiro, V. Potocek, A. Gabris, I. Jex, and Ch. Silberhorn, 
{\it Decoherence and Disorder in Quantum Walks: From Ballistic Spread to Localization},
Phys. Rev. Lett. {\bf 106}, 180403 (2011).
\bibitem{A31}
R. Zhang, H. Qin, B. Tang, and P. Xue, {\it  Disorder and decoherence in coined quantum walks},
Chinese Phys. B {\bf 22}, 110312 (2013).
\bibitem{A31b}
A. Alberti, W. Alt, R. Werner, and D. Meschede,
{\it Decoherence models for discrete-time quantum
walks and their application to neutral atom
experiments}, New J. Phys. 16 123052 (2014).
\bibitem{A32}
F. Caruso, A. Crespi, A.G. Ciriolo, F. Sciarrino, and R. Osellame, {
\it Fast escape of a quantum walker from an integrated photonic maze}, 
Nature Commun. {\bf 7}, 11682 (2016).
\bibitem{A33}
D.N. Biggerstaff, R. Heilmann, A.A. Zecevik, M. Gr\~afe, M.A. Broome, A. Fedrizzi, S. Nolte, A. Szameit, A.G. White, and I. Kassal, {\it Enhancing coherent transport in a photonic network using controllable decoherence},
Nature Commun. {\bf 7}, 11282 (2016).
\bibitem{A34}
G. Bressanini, C. Benedetti, and M.G.A. Paris,
{\it Decoherence and classicalization of continuous-time
quantum walks on graphs}, Quantum Inf. Process. {\bf 21}, 37 (2022).
\bibitem{A34b}
J. Moreno, A. Pendse, and A. Eisfeld, {\it Unraveling of the Lindblad equation
of $N$ coupled oscillators into $N$ independent ones}, Appl. Phys. Lett. {\bf 124},
161110 (2024).
\bibitem{A35}
S. Longhi, {\it 
Incoherent non-Hermitian skin effect in photonic quantum walks},
Light: Sci. \& Appl. {\bf 13}, 95 (2024).
\bibitem{A36}
S. Longhi, {\it Photonic Mpemba effect}, Opt. Lett. {\bf 49}, 5188 (2024).

\bibitem{doubly}
R.B. Bapat and T.E.S. Raghavan,  {\it Doubly stochastic matrices}, in: {\it Nonnegative Matrices and Applications}, Encyclopedia of Mathematics and its Applications, pp. 59-114 (Cambridge University Press, UK, 1997).



\bibitem{chiral}
D. Lu, J.D. Biamonte, J. Li, H. Li, T.H. Johnson, V. Bergholm, M. Faccin, Z. Zimboras, R. Laflamme, J. Baugh, and S. Lloyd, {\it Chiral quantum walks},
Phys. Rev. A {\bf 93}, 042302 (2016).
\bibitem{A37}
M. Casanellas, J. Fernandez-Sanchez  and J.i Roca-Lacostena,
{\it The embedding problem for Markov matrices},
Publ. Mat. {\bf 67}, 411  (2023).


\bibitem{caz1}
J. Kempe, {\it Quantum random walks: An introductory overview}, Contemp. Phys. {\bf 44}, 307 (2003).
\bibitem{caz2}
S.E. Venegas-Andraca, {\it Quantum walks: a comprehensive review}, Quantum Inf. Process {\bf 11}, 1015 (2012).
\bibitem{caz3}
D. Reitzner, D. Nagaj, and V. Buzek, {\it Quantum Walks}, Acta Physica Slovaca 
{\bf 61}, 603 (2011).


\bibitem{trap1}
H. Schmitz, R. Matjeschk, C. Schneider, J. Glueckert, M. Enderlein, T. Huber, and T. Schaetz, {\it Quantum Walk of a Trapped Ion in Phase Space}, Phys. Rev. Lett. {\bf 103}, 090504 (2009).
\bibitem{trap2}
M. Karski, L. F\"orster, J.-M. Choi, A. Steffen, W. Alt, D. Meschede, and A. Widera, {\it Quantum walk in position space with single optically trapped atoms}, Science {\bf 325}, 174 (2009).
\bibitem{trap3}
F. Z\"ahringer, G. Kirchmair, R. Gerritsma, E. Solano, R. Blatt, and C. F. Roos, {\it Realization of a Quantum Walk with One and Two Trapped Ions}, Phys. Rev. Lett. {\bf 104}, 100503 (2010).





\bibitem{cold1}
S. Dadras, A. Gresch, C. Groiseau, S. Wimberger, and G. S. Summy, {\it Quantum Walk in Momentum Space with a Bose-Einstein Condensate}, Phys. Rev. Lett. {\bf 121}, 070402 (2018).


\bibitem{phot1}
A. Schreiber, K.N. Cassemiro, V. Potocek, A. Gabris, P.J. Mosley, E. Andersson, I. Jex, and Ch. Silberhorn,
{\it Photons Walking the Line: A Quantum Walk with Adjustable Coin Operations}, Phys. Rev. Lett. {\bf 104}, 050502 (2010).
\bibitem{phot2}
F. Cardano, A. D'Errico, A. Dauphin, M. Maffei, B. Piccirillo, C. de Lisio, G. De Filippis, V. Cataudella, E. Santamato, L. Marrucci,  M. Lewenstein, and P. Massignan, {\it Detection of Zak phases and topological invariants in a chiral quantum walk of twisted photons}, Nat. Commun. {\bf 8}, 15516 (2017).
\bibitem{phot3}
X. Zhan, L. Xiao, Z. Bian, K. Wang, X. Qiu, B.C. Sanders, W. Yi, and P. Xue,
{\it Detecting Topological Invariants in Nonunitary Discrete-Time Quantum Walks},
Phys. Rev. Lett. {\bf 119}, 130501 (2017).
\bibitem{photo3b}
L. Xiao, X. Zhan, Z.H. Bian, K.K. Wang, X. Zhang, X.P. Wang, J. Li, K. Mochizuki, D. Kim, N. Kawakami, W. Yi, H. Obuse, B.C. Sanders, and P. Xue, {\it Observation of topological edge states in parity-time-symmetric quantum walks}, Nature Phys. {\bf 13}, 1117 (2017).

\bibitem{photo3c}
S. Longhi, {\it Unraveling the non-Hermitian skin effect in dissipative systems},
Phys. Rev. B {\bf 102}, 201103(R) (2020).

\bibitem{phot4}
P. Xue, Q. Lin, K. Wang, L. Xiao, S. Longhi, and W. Yi, {\it Self acceleration from spectral geometry in dissipative quantum-walk dynamics}, 
Nature Commun. {\bf 15}, 4381 (2024). 
\bibitem{phot5}
 L. Neves and G. Puentes,
{\it Photonic Discrete-time Quantum Walks and Applications}, Entropy {\bf 20}, 731 (2018).








\bibitem{A38}
H. Obuse and N. Kawakami, {\it Topological phases and delocalization of quantum walks in random environments}, Phys. Rev. B {\bf 84}, 195139 (2011).
\bibitem{A39}
J.K. Asbot, {\it Symmetries, topological phases, and bound states in the one-dimensional quantum walk}, Phys. Rev. B {\bf 86}, 195414 (2012).
\bibitem{A39b}
J.K. Asboth and H. Obuse, {\it Bulk-boundary correspondence for chiral symmetric quantum walks}
Phys. Rev. B {\bf 88}, 121406 (2013).

\bibitem{figo}
Y. Ren, R. Zhao, K. Ye, L. Zhang, H. Chen, H. Xue, and Y. Yang, {\it Spacetime-disorder-induced localization of light in non-Hermitian quasicrystals}, arXiv:2505.04205  (2025).




\end{thebibliography}

\appendix

\renewcommand{\thesection}{\Alph{section}}
\renewcommand{\thefigure}{A\arabic{figure}}
\renewcommand{\thetable}{A\Roman{table}}
\setcounter{figure}{0}
\renewcommand{\theequation}{A\arabic{equation}}
\setcounter{equation}{0}

\textcolor{black}{
\section{Phase transitions in the full quantum regime}
In the main text, we discussed the occurrence of either first- or second-order dynamical phase transitions in the classical limit $q=1$, where dephasing is applied at each time step and the quantum master equation~(4) reduces to a discrete-time classical Markov process [Eq.~(5)] for the diagonal elements (populations) of the density matrix. 
In this Appendix, we consider the full quantum regime $q<1$, where coherences are not fully damped at each time step. The parameter $q$ basically measures the damping rate of coherences in the dynamics, with $q=1$ corresponding to fully incoherent dynamics (classical limit) and $q=0$ to fully coherent (Hamiltonian) dynamics.  In this Appendix we show that the emergence of first- or second-order dynamical phase transitions, found in the classical (fully incoherent) regime $q=1$, persists in the quantum regime, provided that the damping rate $q$ of coherences is not close to zero.\\ 
In the full quantum regime $0<q<1$, in addition to the diagonal elements $\rho_{n,n}^{(t)}= \langle n | \rho^{(t)} | n \rangle$ (populations), one must also consider the relaxation of the off-diagonal elements $\rho_{n,m}^{(t)}= \langle n | \rho^{(t)} | m \rangle$ with $n \neq m$ (coherences). 
From the quantum master equation~(4), one readily obtains the following map describing the discrete-time evolution of the density matrix elements:
\begin{equation}
\rho_{n,m}^{(t+ \tau)} = (1-q) \sum_{l,r=1}^{N} U_{n,l} U_{m,r}^* \rho_{l,r}^{(t)} + q \delta_{n,m}  \sum_{l,r=1}^{N} U_{n,l} U_{n,r}^* \rho_{l,r}^{(t)},
\label{A1}
\end{equation}
where $U_{n,m}$ are the elements of the one-step coherent evolution propagator $U = \exp(-i H \tau)$. This equation can be formally written as $\rho^{(t+\tau)} = \exp(-\mathcal{L}) \rho^{(t)}$, where the Liouvillian superoperator $\mathcal{L}$ is represented by a $N^2 \times N^2$ matrix.
Clearly, in the classical limit $q=1$, Eq.~(\ref{A1}) implies that all coherences $\rho_{n,m}^{(t)}$ vanish, and the evolution equation for the diagonal elements $P_n^{(t)} = \rho_{n,n}^{(t)}$ reduces to a classical Markov process [Eq.~(5) in the main text] with a $N \times N$ Markov transition matrix $Q$ given by Eq.~(6).
In the full quantum regime $q<1$, it can be readily shown that the stationary state $\rho_s$ of Eq.~(\ref{A1}), i.e. $\exp(- \mathcal{L}) \rho_s=\rho_s$, is diagonal and coincides with the stationary classical distribution $\pi$, i.e., $(\rho_s)_{n,m} = 0$ for $n \neq m$ and $(\rho_s)_{n,n} = \pi_n = 1/N$. This stationary solution is independent of the system parameter $\beta$, and we assume that it is the unique stationary state of the quantum master equation.
The relaxation dynamics toward the non-equilibrium steady state $\rho_s$ is governed by the $N^2$ eigenvalues $\lambda_\alpha$ (Floquet exponents) of the Liouvillian superoperator $\mathcal{L}$, ordered such that $\lambda_1 = 0 < \text{Re}(\lambda_2) \leq \text{Re}(\lambda_3) \leq \cdots \leq \text{Re}(\lambda_{N^2})$, and by their corresponding right and left eigenvectors, $r^{(\alpha)}$ and $l^{(\alpha)}$. As in the classical limit $q=1$, the zero eigenvalue $\lambda_1 = 0$ corresponds to the stationary state $\rho_s$.\\
As the control parameter $\beta$ is varied, a dynamical quantum phase transition occurs when there is a crossing between the real parts of the second ($\lambda_2$) and third ($\lambda_3$) eigenvalues at some critical value $\beta = \beta_c$, indicating a change in the dominant relaxation dynamics. The phase transition is of first order when the lowest-decaying eigenvector undergoes a discontinuous jump as $\beta$ is varied across the critical value $\beta_c$, i.e., the eigenvalue crossing does not correspond to an exceptional point with simultaneous coalescence of eigenvectors. Conversely, the phase transition is of second order when the eigenvalue crossing corresponds to a simultaneous coalescence of the eigenvectors, i.e., an exceptional point. To characterize the nature of the phase transition, one can calculate the behavior versus $\beta$ of the scalar product (eigenvector overlapping)
\begin{equation}
g(\beta)=\frac{| \langle r^{(3)} | r^{(2)} \rangle | } {\sqrt{ \langle r^{(2)} | r^{(2)} \rangle \langle r^{(3)} | r^{(3)} \rangle  }}
\end{equation}
near the crossing point $\beta_c$.
Clearly, $0 \leq g \leq 1$ and $g=1$ when there is eigenvector coalescence ($r^{(2)}=r^{(3)}$): hence a second-order phase transition is characterized by the condition $g(\beta_c)=1$, while $g(\beta_c)<1$ indicates a first-order phase transition.\\
To illustrate the onset of first- and second-order phase transitions in the full quantum regime $q<1$, we consider the minimal quantum network architecture comprising $N=3$ sites, shown in Fig.~1(c). The coherent Hamiltonian $H$ of the network is given by Eq.~(9), and it exhibits time-reversal symmetry for a vanishing gauge phase $\phi = 0$. As a control parameter $\beta$, we take the coherent time interval $\tau$ normalized to the hopping rate $J_1$, i.e., $\beta = J_1 \tau$. The Liouvillian superoperator $\mathcal{L}$ is described by a $9 \times 9$ matrix, and its eigenvalues and corresponding left/right eigenvectors can be readily computed numerically. The classical limit $q=1$ is discussed in the main text and illustrated in Figs.~2 and 3.
In the full quantum regime $q<1$, we observe a similar qualitative behavior to that of the classical case $q=1$ provided that $q$ is sufficiently close to one: a first-order phase transition is found when the Hamiltonian $H$ exhibits time-reversal symmetry ($\phi = 0$), whereas second-order phase transitions arise when time-reversal symmetry is broken. This is illustrated in Figs.~A1 and A2. 
Figure~A1 shows the typical behavior of the real parts of the two eigenvalues $\lambda_2$ and $\lambda_3$ (decay rates)  versus $\beta$, and of the eigenvector overlapping $g(\beta)$, for a few decreasing values of $q$. The parameter values are $J_1 = J_2 = 1$, $J_3 = 0.5$, and $\phi = 0$, corresponding to unbroken time-reversal symmetry. As one can see, for $q$ sufficiently close to one an eigenvalue crossing occurs at a critical value $\beta_c$, without coalescence of the corresponding eigenvectors -- characteristic of a first-order phase transition. The critical value $\beta_c$ decreases as $q$ is reduced, indicating that in the full quantum regime $q<1$, the phase transition occurs at a smaller $\beta_c$ than in the classical limit. As $q$ is further reduced to reach the critical value $q=q_c \simeq 0.23$, the eigenvalue crossing -- and thus the phase transition-- disappears (the two eigenvalues $\lambda_2$ and $\lambda_3$ have the same decay rate), as shown in Fig.A1.\\   
Figure~A2 displays of the behavior of real parts of the two eigenvalues $\lambda_2$ and $\lambda_3$ (decay rates)  versus $\beta$, and of the eigenvector overlapping parameter $g(\beta)$,  for the same parameters as in Fig.~A1, except with $\phi = \pi/3$, corresponding to broken time-reversal symmetry. In this case, a second-order phase transition is observed when $q$ is close to one, characterized by an exceptional point-- i.e., the eigenvalue crossing is accompanied by the coalescence of the corresponding eigenvectors. As in the previous case, the critical value $\beta_c$ at which the level crossing occurs is reduced in the quantum regime $q<1$ compared to the classical limit. Below a critical value $q=q_c \simeq 0.356$, the phase transition is not found anymore.  
\begin{figure}
\includegraphics[width=19 cm]{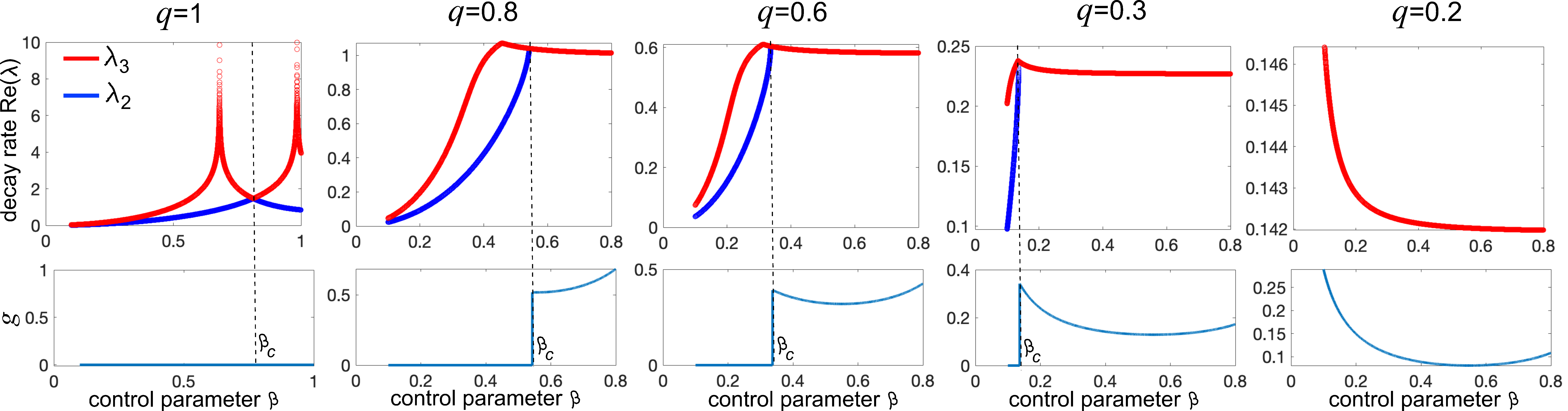}
\caption{ \textcolor{black} {Effect of the dephasing probability $q$ on first-order phase transition for the minimal network of Fig.1(c) with the same parameter values as in Fig.2 and for decreasing values of $q$. The upper panels show the behavior of the decay rates of second and third eigenvectors of the Liouvillian superoperator $\mathcal{L}$ versus the control parameter $\beta$, whereas the lower panels depict the corresponding behavior of the eigenmode overlapping parameter $g=g(\beta)$. The first column, $q=1$, corresponds to the classical limit discussed in the main text (Fig.2). As $q$ is decreased, the critical value $\beta_c$ of eigenvalue crossing decreases. No eigenvalue crossing occurs for $q<q_c \simeq 0.23$.}}
\end{figure}
\begin{figure}
\includegraphics[width=19 cm]{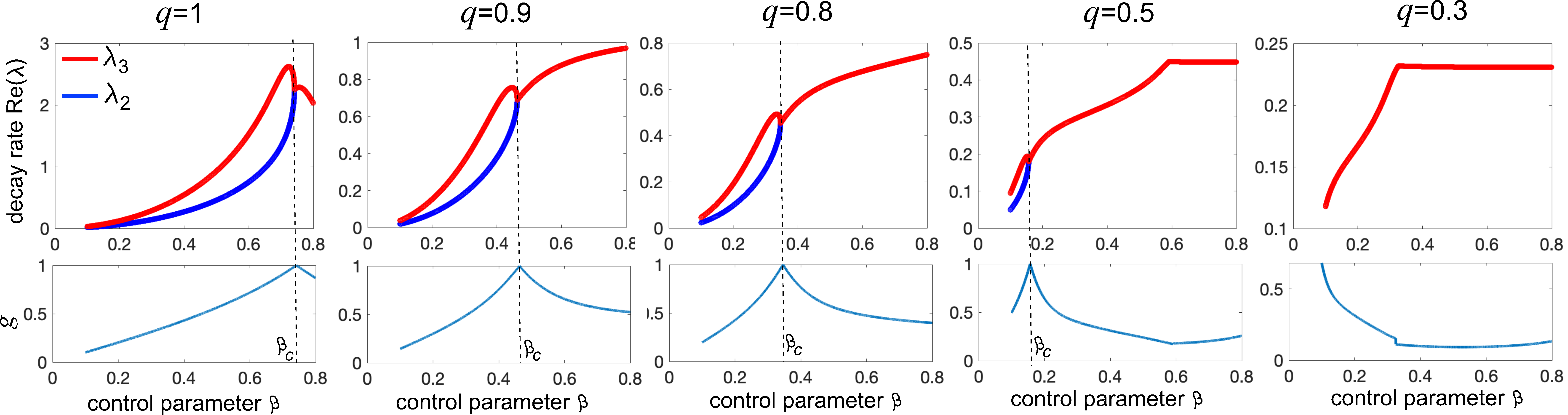}
\caption{ \textcolor{black} {Effect of the dephasing probability $q$ on second-order phase transition for the minimal network of Fig.1(c) with the same parameter values as in Fig.3 and for decreasing values of $q$. The upper panels show the behavior of the decay rates of second and third eigenvectors of the Liouvillian superoperator $\mathcal{L}$ versus the control parameter $\beta$, whereas the lower panels depict the corresponding behavior of the overlap $g=g(\beta)$ of the two eigenvectors. The first column, $q=1$, corresponds to the classical limit discussed in the main text (Fig.3). As $q$ is decreased, the critical value $\beta_c$ of eigenvalue crossing decreases. No eigenvalue crossing occurs for $q<q_c \simeq 0.356$.}}
\end{figure}
 }

\end{document}